# Hanle-effect measurements of spin injection from $Mn_5Ge_3C_{0.8}/Al_2O_3$-contacts into degenerately doped Ge channels on Si

Inga Anita Fischer,[1,*] Li-Te Chang,[2,*] Christoph Sürgers,[3] Erlend Rolseth,[1] Sebastian Reiter,[1] Stefan Stefanov,[4] Stefano Chiussi,[4] Jianshi Tang,[2] Kang L. Wang,[2] and Jörg Schulze[1]

[1] Institut für Halbleitertechnik (IHT), Universität Stuttgart, Pfaffenwaldring 47, Stuttgart, 70569, Germany

[2] Device Research Laboratory, Department of Electrical Engineering, University of California, Los Angeles, CA 90095, USA

[3] Physikalisches Institut, Karlsruhe Institute of Technology, Wolfgang-Gaede-Str. 1, Karlsruhe, 76131, Germany

[4] Dpto. Física Aplicada, E. I. Industrial, Universidad de Vigo, Vigo, Spain

We report electrical spin injection and detection in degenerately doped n-type Ge channels using $Mn_5Ge_3C_{0.8}/Al_2O_3/n^+$-Ge tunneling contacts for spin injection and detection. The whole structure is integrated on a Si wafer for complementary metal-oxide-semiconductor (CMOS) compatibility. From three-terminal Hanle-effect measurements we observe a spin accumulation up to 10 K. The spin lifetime is extracted to be 38 ps at $T = 4$ K using Lorentzian fitting, and the spin diffusion length is estimated to be 367 nm due to the high diffusion coefficient of the highly doped Ge channel.

During the last decades, a continuous downscaling of CMOS field-effect-transistor (FET) dimensions has lead to an increase in device performance and efficiency.[1] With gate lengths of currently 14 nm in production, this downscaling is approaching its end, and alternative device concepts that make use of state variables other than electronic charge are being increasingly investigated as potential candidates for logic devices. Among these device concepts are spintronic devices, in which the electron spin is used in addition to (or instead of) the electron charge for information processing. The current spin polarization in a semiconductor (SC) can be defined as $P_{J,sc} = (J_\uparrow - J_\downarrow)/(J_\uparrow + J_\downarrow)$, where $J_\uparrow$ and $J_\downarrow$ are the current densities for spin-up and spin-down electrons, respectively, and it describes the asymmetry of majority and minority spin currents in the semiconductor. The prototypical spin-polarized FET was proposed by Datta and Das in 1990[2], which was followed by numerous other concepts for spintronic devices.[3-7] In these concepts, two achievements are necessary for the functioning of the device: one is the efficient injection and detection of spin-polarized currents in the semiconductor channel, the other is the effective manipulation of spin-polarized currents. A lot of progress has been made to establish spintronics in the industrially most relevant semiconductor Si.[8-11] More recently, Ge has been investigated as another promising candidate for CMOS-compatible spintronics because of its high electron mobility that is only weakly dependent on dopant concentration.

To demonstrate spin injection into a semiconductor channel, the conductivity mismatch problem[12,13] can be solved by inserting a thin oxide layer for electron tunneling between the ferromagnetic metal and the semiconductor channel. Spin injection was demonstrated in Ge by using tunnel contacts[14-16] and Schottky contacts.[17] $Mn_5Ge_3$ is a ferromagnet (FM) with a Curie temperature $T_C \approx 300$ K[18,22] that can be enhanced by C-doping up to $T_C = 450$ K.[19-21] C-doped $Mn_5Ge_3C_x$ has a negligible conductivity mismatch with highly doped Ge[22], which makes it particularly interesting as a contact material for spin injection into Ge. Furthermore, $Mn_5Ge_3C_x$ can be grown epitaxially on Ge (111)[21], which could help to avoid the formation of interface defects and hence improve the spin injection/detection efficiency.

In this work, we investigate $Mn_5Ge_3C_{0.8}$ contacts sputtered on degenerately Sb-doped $n$-Ge (111)-channels ($N_D = 1\times10^{20}$ cm$^{-3}$) grown on a Si (111) wafer. We demonstrate spin injection using three-terminal Hanle-effect measurements under various bias conditions between 1.9 K and 10 K, and estimate the spin lifetime ($\tau_s$) as well as the spin diffusion length ($\lambda$) from Lorentzian fits to the data. On the other hand, because spin injection into Ge has previously been demonstrated in devices fabricated on Ge substrates, the integration of such Ge channels on Si substrates holds great potential for future CMOS-compatible SiGe-based spintronic devices.

Doped semiconductor layers were grown by solid-source molecular beam epitaxy (MBE). First, a Ge virtual substrate with a thickness of 50 nm was deposited on a Si (111) substrate. It accommodates the lattice mismatch between the Si substrate and the final Ge layer, and thus enables the subsequent growth of high-quality Ge layers. This was followed by the growth of a 300 nm $n$-Ge buffer layer with $N_D = 1\times10^{18}$ cm$^{-3}$. Finally, a thin layer (40 nm) of degenerately doped $n^+$-Ge ($N_D = 1\times10^{20}$ cm$^{-3}$) was deposited as highly conductive channel layer. The device fabrication process started with structuring of the mesa by reactive ion etching (RIE). A SiO$_2$ layer (100 nm) was deposited by plasma-enhanced chemical vapor deposition (PECVD) to passivate the surface. Contact holes (size 98 × 198 μm$^2$) were defined by photolithography and structured by RIE as well as wet chemical etching. Prior to the deposition of Mn$_5$Ge$_3$C$_{0.8}$, a thin Al$_2$O$_3$ layer (2 nm) was deposited by plasma-enhanced atomic layer deposition (PE-ALD) in order to prevent material intermixing. The 40-nm Mn$_5$Ge$_3$C$_{0.8}$ contacts were deposited by simultaneous dc- and rf-magnetron sputtering from elemental targets of Mn, Ge, and C in a high-vacuum system under Ar atmosphere at a substrate temperature $T_s = 400$ °C. The Mn$_5$Ge$_3$C$_{0.8}$ films were capped *in-situ* with 40 nm of Al at $T_s = 40$ °C. After a 5 s dip in buffered hydrofluoric acid (BHF), another 300-400 nm of Al were then deposited *ex-situ* by thermal evaporation as the final metallization layer. The Al/Mn$_5$Ge$_3$C$_{0.8}$-metallization was patterned by photolithography and structured in one etching step. A reference sample without the ferromagnetic Mn$_5$Ge$_3$C$_{0.8}$ layer was also fabricated.

The three-terminal measurements setup and the cross section of the active contact area are shown schematically in Fig. 1 (a) and (b). The temperature-dependent three-terminal Hanle-effect measurements were carried out in a Quantum Design physical property measurements system (PPMS). As shown in Fig. 1 (a), a DC current was applied between contact 2 and 3 with a Keithley 6221 DC/AC current source, and the voltage difference $V_{12}$ was measured between contact 2 and 1 with a Keithley 2182 nanovoltmeter. Temperature-dependent *I-V* measurements between two adjacent Mn$_5$Ge$_3$C$_{0.8}$/Al$_2$O$_3$/$n^+$-Ge tunnel contacts (contact 1 and 2) are shown in Fig. 1 (c), and the temperature dependence of the normalized zero-bias resistance $R_0(T)/R_0(300\ K)$ is shown in the inset. The moderate temperature dependence suggests the total current consists of both tunneling and leakage components for a large active contact (98 × 198 μm$^2$). Fig. 1 (d) shows a simulated conduction band diagram of the Mn$_5$Ge$_3$C$_{0.8}$/Al$_2$O$_3$/$n^+$-Ge tunnel contact at 10 K under zero bias that is obtained by solving the Poisson and Schrödinger equations self-consistently in one dimension.[23] The sharp tunneling barrier and a well-defined potential reservoir in the conduction band of $n^+$-Ge near the Al$_2$O$_3$/$n^+$-Ge interface are two essential features for efficient spin injection and detection.[24]

High-resolution transmission electron microscopy (HRTEM) was used to characterize the Mn$_5$Ge$_3$C$_{0.8}$/Al$_2$O$_3$/$n^+$-Ge junction. Material intermixing can occur at the Mn$_5$Ge$_3$C$_{0.8}$/$n^+$-Ge interface if Mn$_5$Ge$_3$C$_{0.8}$ is directly sputtered on top of a Ge layer, as can be seen in Fig. 2 (a). To ensure material separation and enhance spin polarization, an Al$_2$O$_3$ tunneling oxide was inserted at the interface.[25] Atomic-force microscopy (AFM) on a reference sample without the Al/Mn$_5$Ge$_3$C$_{0.8}$-metallization was used to measure the roughness of the Al$_2$O$_3$ surface in order to obtain information on the Mn$_5$Ge$_3$C$_{0.8}$/Al$_2$O$_3$ interface. Fig. 2 (c) shows an AFM image of the surface of the Al$_2$O$_3$ tunneling oxide with a square area of $500 \times 500$ nm$^2$, and Fig. 2 (d) shows a cross-sectional height profile along the corresponding double heads arrow. We obtained a root-mean-square (rms) roughness of 0.2 nm with a correlation length of 30 nm, which is comparable to the results obtained in Ref. 26 and suggests a reduced spin accumulation as well as spin lifetime from the effects of local magnetostatic fields.

Fig.s 3 (a) and (b) show voltage signals $\Delta V$ measured while a magnetic field was applied perpendicularly ($B^\perp$) to the Mn$_5$Ge$_3$C$_{0.8}$/Al$_2$O$_3$/$n^+$-Ge-interface or parallel ($B^{||}$) to the direction of the long axis of the Mn$_5$Ge$_3$C$_{0.8}$ electrodes, respectively. The voltage signal $\Delta V(B)$ is obtained by subtracting a background voltage $V_{back} \sim B^2$ from the raw data, i.e., $\Delta V(B) = V_{12}(B) - V_{back}(B)$. The measurements were performed at 4 K under various spin injection and extraction

conditions. Recently, Song and Dery argued that such Lorentzian-shaped magnetoresistance (MR) signals could qualitatively and quantitatively be explained by resonant tunneling through localized impurity states within the tunnel barrier[27]. Impurity-assisted tunneling MR also provides an alternative interpretation of recent experiments[28,29]. Therefore, it is necessary to rule out the barrier as the origin of the measured spin signals. Figure 3 (c) compares the measurements performed on a $Mn_5Ge_3C_{0.8}/Al_2O_3/n^+$-Ge sample and a $Al/Al_2O_3/n^+$-Ge reference sample under different field orientations ($B^\perp$ or $B^{||}$) at $I = -0.5$ mA. A very small dip in the $\Delta V(B)$ curves of the reference sample around zero magnetic field can be observed. While we cannot completely exclude a contribution from resonant tunneling through localized impurity states in the $Al_2O_3$ barrier, that effect alone is clearly unable to explain the voltage signals from the $Mn_5Ge_3C_{0.8}/Al_2O_3/n^+$-Ge sample. We, therefore, attribute the observed $\Delta V(B)$ behavior in our $Mn_5Ge_3C_{0.8}/Al_2O_3/n^+$-Ge sample to Hanle signals (in $B^\perp$) and inverted Hanle signals (in $B^{||}$) arising from spin accumulation in the semiconductor channel[30].

The Hanle signal indicates a reduction of spin polarization as a result of spin precession induced by a perpendicular magnetic field, strongly indicating that spin accumulation is present in the Ge channel.[26] On the other hand, the appearance of the inverted Hanle effect is a consequence of spatially inhomogeneous magnetostatic stray-fields that arise from the finite roughness of the $Mn_5Ge_3C_{0.8}/Al_2O_3$-interface, which enhance incoherent spin precession and reduce spin polarization. In the inverted Hanle effect, an in-plane magnetic field aligns the conduction electron spins in the semiconductor channel, and therefore reduces the diminishing effect of spin precession at the interface and restores spin polarization.[26]

The qualitatively most notable features in our measurements are the signs of the Hanle and inverted Hanle signals under spin injection ($I < 0$) and extraction ($I > 0$) conditions. The measured Hanle signal $\Delta V$ is determined by $\Delta V \propto P_i \times P_d \times I$, where $P_i$ and $P_d$ are the spin polarizations for spin injector and spin detector, respectively. In the present case, $P_i$ and $P_d$ have the same sign, so the Hanle signal $\Delta V$ should be positive for spin extraction conditions and negative for spin injection, but the present samples show the opposite trends in Fig. 3 (a). This is also seen in Fig. 3 (d), where the maximum signal $\Delta V$ at $B = 0$ is plotted for the Hanle and inverted Hanle effect. Sign inversion of the Hanle signal $\Delta V$ has been observed in three-terminal devices for different electrode materials using Schottky as well as tunneling contacts, and the sign has been reported to change under different applied bias and temperature.[31-33]

Two main mechanisms have been discussed for the inversion of spin polarization at an FM/oxide/SC interface. Those are tunneling via interfacial resonant states[34-36] and tunneling from bound states into a heavily doped semiconductor surface layer, from which minority-spin states are preferentially extracted[24,37]. Fig. 3 (d) also shows that $\Delta V(B = 0)$ of the Hanle and inverted Hanle signals change sign at different values of the applied bias. In our samples the $Mn_5Ge_3C_{0.8}$ contact is polycrystalline and the orientation of the magnetization is not perfectly aligned in all crystal grains. As a result, even a magnetic field aligned parallel to the $Mn_5Ge_3C_{0.8}/Al_2O_3/n^+$-Ge-interface can induce spin precession and, as a consequence, reduce spin accumulation. Future experiments are however required to explore the physical origins of the sign change of the Hanle as well as the inverted Hanle signal. In particular, the influence of a variation in the thickness of the highly doped surface Ge layer should be investigated in order to probe the influence of bound states in the surface layer, and spin injection from polycrystalline and single-crystalline $Mn_5Ge_3C_x$ contacts could be compared.

Fig. 4 (a) shows the Hanle signal and the related Lorentzian fit at $T = 2$ K. The amplitude of the Hanle signal decreases with increasing temperature, as shown in Fig. 4 (b), and no Hanle signal is observed for temperatures above 10 K. We extract the spin lifetime $\tau_s$ by fitting the data with Lorentzian curves according to:

$$\Delta V = \frac{\Delta V_0}{1+(\omega_L \tau_s)^2} = \frac{\Delta V_0}{1+\left(B\frac{\tau_s g \mu_B}{\hbar}\right)^2} = \frac{\Delta V_0}{1+\left(\frac{B}{\Delta B}\right)^2}$$

where $\omega_L = g\mu_B B/\hbar$ is the Larmor frequency, $\mu_B$ is the Bohr magneton, $g = 1.6$ is the Landé factor[38], and $\hbar = h/2\pi$ ($h$: Planck constant). Spin accumulation is reduced by 50% at the point where $B = \Delta B = \hbar / g\mu_B \tau_s$, i.e., we obtain the spin lifetime from the half width at half maximum $\Delta B$.

The results for our sample are shown in Fig. 4 (c). We find that the spin lifetime of 38 ps at 10 K obtained for our samples is comparable to the 35 ps at 10 K reported in Ref. 39, where spin injection from Co/permalloy into n-type Ge via a $Al_2O_3$ tunneling barrier was investigated in a three terminal geometry. As mentioned above, magnetostatic stray-fields due to the roughness of the ferromagnetic electrode at the interface could induce artificial broadening of the Hanle peaks.[26] Therefore, the spin lifetime extracted from the width of the Hanle signal has to be taken as a lower bound.

Generally, Elliott-Yafet spin relaxation is expected to dominate for conduction electrons in *n*-type Ge.[14] In a degenerately doped system, the Elliott-Yafet spin relaxation time $\tau_{EY}$ is proportional to the momentum relaxation $\tau_p$.[40] For our samples, the temperature dependence of $\tau_p$ is weak over the investigated temperature range ($T < 10$ K) as can be seen in Fig. 4 (c). This is because the Hall mobility $\mu_{el}(T) = q\tau_p(T)/m^*_{e,Ge}$ (*q*: electron charge, $m^*_{e,Ge}$: effective electron mass) is almost constant with $\mu_{el}(T) = 400\ cm^2/V \cdot s$ for $T < 10$ K, as shown in the inset of Fig. 4 (c). Therefore, the weak temperature dependence of spin lifetime is consistent with measurements of the Hall mobility for $T < 10$ K. The experimental observation that spin injection is only possible at very low temperatures can have two reasons. On the one hand, the high concentration of dopants can induce large spin scattering at all temperatures. On the other hand, threading dislocations in the Ge channel can reduce spin lifetimes at all temperatures.

From the spin lifetime we estimate the spin diffusion length $\lambda = \sqrt{D(T)\tau_s}$, where $D(T)$ is the temperature-dependent electron diffusion coefficient. In our degenerately doped samples we can determine *D* according to:[41]

$$D = \frac{2}{3q}(E_F - E_c) \cdot \mu_{el}(T), \qquad (2)$$

where we approximate

$$E_F - E_C = \frac{1}{m^*_{e,Ge}}\left(\frac{3}{2}\frac{\pi^2\hbar^3}{\sqrt{2}}N_D\right)^{2/3} = 142.8\ meV \qquad (3)$$

at $T = 0$, using $N_D = 1\times 10^{20}$ cm$^{-3}$ and $m^*_{e,Ge} = 0.55\ m_{e,0}$. By taking into account the mobility data obtained from Hall measurements on our sample as shown in Fig. 4 (c) (inset), we obtain a temperature-dependent diffusion coefficient $D \sim 37.5$ cm$^2$/s for electrons in the $n^+$-Ge layer and a corresponding spin diffusion length of 397 nm at $T = 2$ K. This would indicate that, even though the spin lifetime is low, we observe spin-diffusion lengths sufficiently large for device applications as a result of the large diffusion coefficient in the Ge channel.

In conclusion, we experimentally observed Hanle signals in a $Mn_5Ge_3C_{0.8}/Al_2O_3/n^+$-Ge-system on a Si (111) wafer, indicating that spin injection can be achieved in degenerately doped Ge channels that are integrated on Si. Those Hanle signals are not observed in samples without the ferromagnetic $Mn_5Ge_3C_{0.8}$ layer. For future spintronic device applications, this provides the possibility of using high mobility Ge channels in a CMOS-compatible process flow. Although the spin lifetimes that we measured are still short (38 ps at 10 K), measurements of the carrier mobility let us conclude that spin transport in these channels is possible over distances larger than 100 nm, i.e. on length scales that are relevant for future device operation. The spin signal is suppressed at temperatures > 10 K. We expect to increase this temperature in future experiments by adjusting the doping concentration and improving the Ge layer quality. Finally, it is desirable to compare the results to data obtained from four-terminal measurements in order to obtain precise information on spin lifetime and spin polarization in the Ge channel.

Acknowledgments

LTC, JT, and KLW acknowledge the support from Western Institution of Nanoelectronics (WIN) and National Science Foundation ECCS 1308358.

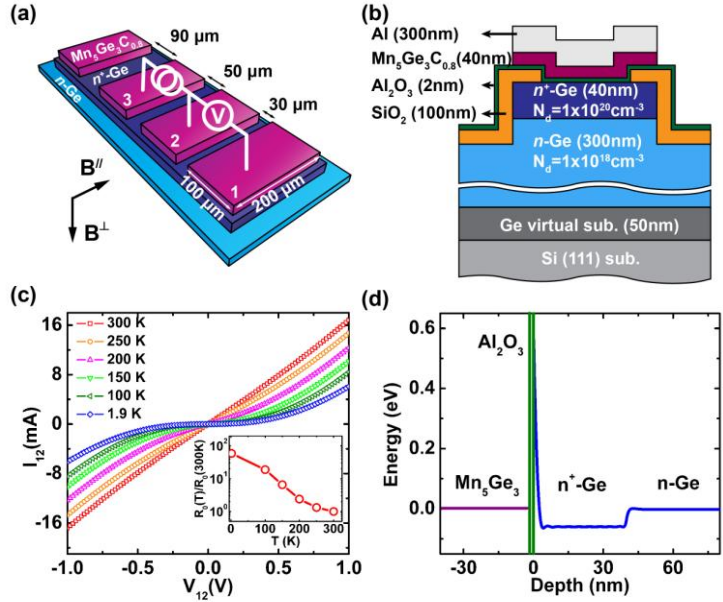

Fig. 1: Device structure and contact properties. (a) Schematic illustration of device structure and measurements setup, where the arrows indicate the magnetic field in the perpendicular ($B^{\perp}$) and parallel ($B^{//}$) direction. (b) Cross section structure of the $Mn_5Ge_3C_{0.8}/Al_2O_3/n^+$-Ge tunneling contact. (c) Temperature dependent *I-V* measurements between contacts 1 and 2. The inset shows the temperature dependence of the normalized zero-biased resistance. (d) Simulated conduction band profile under zero bias at 10 K.

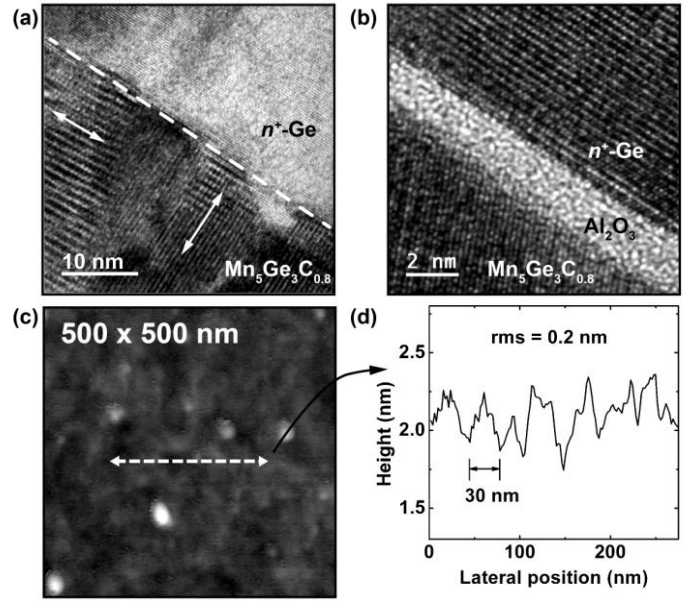

Fig. 2: (a) High-resolution TEM image of a $Mn_5Ge_3C_{0.8}/n^+$-Ge contact. Arrows indicate two different orientations of crystal planes. (b) High-resolution TEM image at the interface of a $Mn_5Ge_3C_{0.8}/Al_2O_3/n^+$-Ge (111) tunneling contact. (c) AFM image of the $Al_2O_3$ surface prior to metal electrode deposition. (d) Cross-sectional height profile along the dashed line in (c).

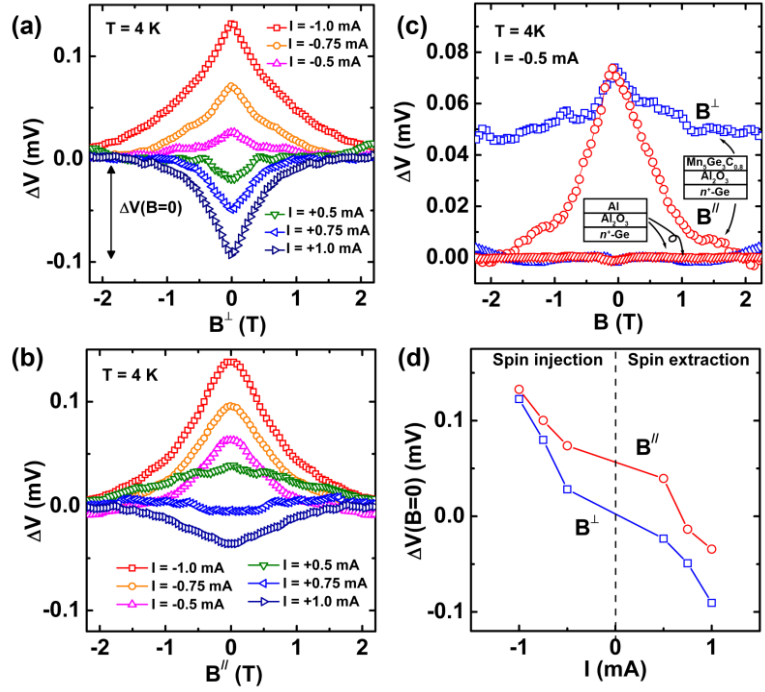

Fig. 3: Voltage signals $\Delta V$ in magnetic fields (a) $B^{\perp}$ and (b) $B^{//}$ at $T = 4$ K for various spin injection and extraction conditions. (c) $\Delta V$ at $T = 4$ K and $I = -0.5$ mA of $Mn_5Ge_3C_{0.8}/Al_2O_3/n^+$-Ge and of the $Al/Al_2O_3/n^+$-Ge reference sample. (d) Peak voltage $\Delta V (B = 0)$ as a function of applied current.

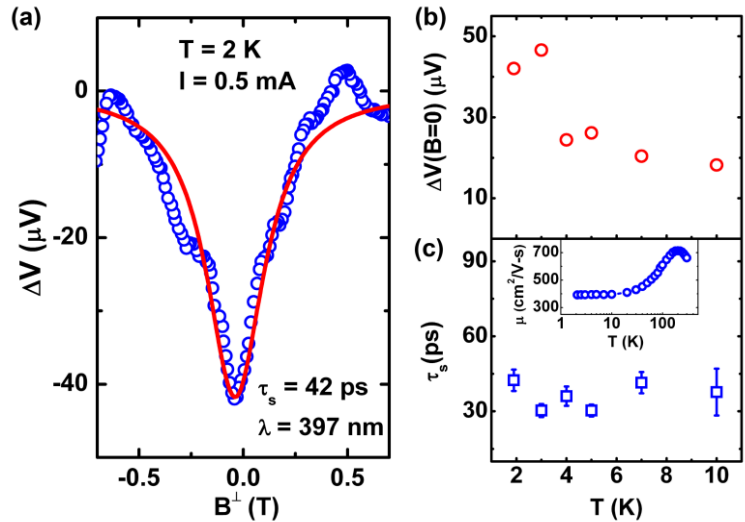

Fig. 4: (a) Hanle signal at $T = 2$ K and $I = 0.5$ mA with Lorentzian fit (red line) to the data. (b) Peak voltage as a function of temperature. (c) Spin lifetime extracted from Lorentzian fitting for temperatures between 2 K and 10 K. The inset shows the temperature-dependent Hall mobility of the same $n^+$-Ge sample in a semi-logarithmic plot.